\documentclass[12pt]{article}
\begin{document}
\title{FUZZY SPACETIME ISSUES}
\author{B.G. Sidharth\\
International Institute for Applicable Mathematics \& Information Sciences\\
Hyderabad (India) \& Udine (Italy)\\
B.M. Birla Science Centre, Adarsh Nagar, Hyderabad - 500 063 (India)}
\date{}
\maketitle
\begin{abstract}
We argue that the Path Integral formulation of Feynman can be reconciled via a Planck scale underpinning for spacetime, with fuzzy spacetime considerations.
\end{abstract}
\section{The Path Integral Formulation}
We first argue that the Feynman Path Integral formulation essentially thrown up fuzzy spacetime. To recapitulate \cite{fh,nottale,it}, if a path is given by
$$x = x(t)$$
Then the probability amplitude is given by
$$\phi (x) = e^{\imath \int^{t_s}_{t_1}L(x, \dot{x})dt}$$
So that the total probability amplitude is given by
$$\sum_{x(t)} \phi (x) = \sum e^{\imath \int^{t_2}_{t_1}L(x,\dot{x})dt} \equiv \sum e^{\frac{\imath}{\hbar}S}$$
In the Feynman analysis, the path
$$x = \bar{x} (t)$$
appears as the actual path for which the action is stationery. For paths very close to this, there is constructive interference, whereas for paths away from this the interference is destructive.\\

We notice that this is also the formulation of the random phase \cite{huang}. However it is well known that the convergence of the integrals requires the well known Lipshitz condition viz.,
\begin{equation}
\Delta x^2 \approx a \Delta t\label{e1}
\end{equation}
We could say that only those paths satisfying (\ref{e1}) constructively interfere. We would now like to observe that (\ref{e1}) is the well known Brownian or Diffusion Equation. This has been commented upon extensively \cite{heap,cu,uof} and also \cite{rief}. The point is that (\ref{e1}) implies a minimum spacetime cut off, as indeed was noted by Feynman himself \cite{fh}, for if $\Delta t$ could $\to 0$, then the velocity would $\to \infty$.\\
As Feynman and Hibbs put it, ``... these irregularities are such that the ``average'' square velocity does not exist, where we have used the classical analogue in referring to an ``average.''\\
If some average velocity is defined for a short time interval $\Delta t$, as, for example, $[x(t + \Delta t) - x(t)]/\Delta t$, the ``mean'' square value of this is $-\hbar /(\imath m \Delta t)$. That is, the ``mean'' square value of a velocity averaged over a short time interval is finite, but its value becomes larger as the interval becomes shorter.\\
It appears that quantum-mechanical paths are very irregular. However, these irregularities average out over a reasonable length of time to produce a reasonable drift, or ``average'' velocity, although for short intervals of time the ``average'' value of the velocity is very high...''\\
To put it another way we are taking averages over an interval $\Delta t$, within which there are totally random and unphysical processes. It is only after the average is taken, that we recover physical spacetime \cite{upcsf}.\\
As has been argued in detail \cite{cu,uof} this is exactly the situation which we encounter in the Dirac theory of the electron. There we have the unphysical zitterbewegung effects within the Compton time $\Delta t$ and as $\Delta t \to 0$ the velocity of the electron $\to \infty$ exactly as above \cite{dirac}. It is only after averaging over the Compton scale that we recover meaningful physics, that is Hermitian position operators.\\
This  existence of a minimum spacetime scale, it has been argued for quite sometime is the origin of fuzzy spacetime, described by a noncommutative geometry, consistent with Lorentz invariance. This was shown a long time ago by Snyder \cite{snyder}. In this case we have commutative relations like
$$[x_\imath , x_j] = \Theta_{\imath j} O(l^2)$$
\begin{equation}
[x_\imath , p_j] = \tilde{\Theta}_{\imath j} \hbar [1 + O(l^2)]\label{e2}
\end{equation}
It is interesting to note that the momentum position commutation relations lead to the usual Quantum Mechanical commutation relations in the usual (commutative spacetime) if $O(l^2)$ is neglected where $l$ defines the minimum scale. Indeed, as noted elsewhere, we have at the smallest scale, a quantum of area. In other words Snyder's purely classical considerations at a Compton scale lead to Quantum Mechanics.
\section{Zero Point Field}
There has been much work connecting the Zero Point Field (or Dark Energy), not only with the new cosmological scenario of an ever expanding universe, which infact was predicted by the author \cite{ijmpa}, but also in the context of a Planck scale underpinning of the universe--the Planck scale being the special case of the more general Compton scale. It was infact argued that $N' \sim 10^{120}$ Planck oscillators provide an underpinning for the entire universe on the one hand, and on the other, throw up the $N \sim 10^{80}$ elementary particles \cite{psp,psu}.\\
It is interesting to note that the Zero Point Field itself leads to the Compton scale \cite{bgsapeiron}. Interestingly the momentum operators $m\vec{v}$ do not satisfy the Quantum Mechanical commutation relations with the position coordinators. But if we add the electromagnetic momentum due to the background Zero Point Field, then we recover the Quantum Mechanical commutation relations and spin half \cite{sachi}. Thus it appears that Classical Mechanics together with the minimum Compton scale, or alternatively Classical Mechanics together with the Zero Point Field leads to Quantum Theory. Infact several authors like Marshall, Boyer and others had argued for Quantum Mechanics arising from stochastic electrodynamics \cite{depena}.\\
Indeed we could even compute the energy due to the Zero Point Field, which in turn gives rise to the Lorentz force and recover the inertial energy.\\
We now observe that the coherent $N'$ Planck oscillators referred to above could be considered to be a degenerate Bose assembly. In this case as is well known we have $(z \approx 1)$
$$v = \frac{V}{N}$$
(Cf.ref.\cite{huang}). $V$ the volume of the universe $\sim 10^{84}$. Whence
$$v = \frac{V}{N'} \sim 10^{-36}$$
So that the wavelength 
\begin{equation}
\lambda \sim (v)^{1/3} \sim 10^{-12} = l\label{e3}
\end{equation}
What is very interesting is that (\ref{e3}) gives us the Compton length of a typical elementary particle like the pion. So from the Planck oscillators we are able to recover the elementary particles exactly as in the references \cite{psp,psu}.\\
Moreover, let us now consider the distant background assembly which is at nearly the same energy. In this case we have (Cf.ref.\cite{huang})
\begin{equation}
\langle n_{\vec{k}}\rangle = \frac{2}{e^{\beta h w}-1}\label{e4}
\end{equation}
As we have assumed that the photons all have nearly the same energy, we have,
\begin{equation}
\langle n_{\vec {k}}\rangle = \langle n_{\vec{k}'}\rangle \delta (k - k')\label{e5}
\end{equation}
where $\langle n_{\vec{k}'}\rangle$ is given by (\ref{e4}), and $k \equiv |\vec{k}|$. The total number of photons $N$, in the volume $V$ being considered, can be obtained in the usual way,
\begin{equation}
N = \frac{V}{(2\pi)^3}\int^\infty_0 \, dk4\pi k^2\langle n_k\rangle\label{e6}
\end{equation}
where $V$ is large. Inserting (\ref{e5}) in (\ref{e6}) we get,
\begin{equation}
N = \frac{2V}{(2\pi)^3} 4\pi k'^2 [\epsilon^\Theta - 1]^{-1} [k]\, \Theta \equiv \beta h w\label{e7}
\end{equation}
where $[k]$ is a dimensionality constant, introduced to compensate the loss of a factor $k$ in the integral (\ref{e6}), owing to the $\delta$-function in (\ref{e5}).\\
We observe that, $\Theta =hw/KT \approx 1$, since by (\ref{e5}), the photons have the same energy $hw$. We also use,
\begin{equation}
v = \frac{V}{N}, \lambda = \frac{2\pi c}{w} = \frac{2\pi}{k} \, \mbox{and}\, z = \frac{\lambda^3}{v}\label{e8}
\end{equation}
$\lambda$ being the wavelength of the radiation. We now have from (\ref{e7}), using (\ref{e8}),
$$(e - 1) = \frac{vk'^2}{\pi^2} [k] = \frac{8\pi}{k'}\frac{1}{z} [k]$$
At this stage we observe that as $z$ is dimensionless, this equation is perfectly consistent because both sides are dimensionless. Using (\ref{e8}) we get:
\begin{equation}
z = \frac{8\pi}{k'(e-1)} = \frac{4\lambda}{(e-1)} [k]\label{e9}
\end{equation}
From (\ref{e9}) we conclude that, in this case,
\begin{equation}
\lambda = \frac{e-1}{4} = 0.4\label{e10}
\end{equation} 
It must be observed that we consider the degenerate case in all the above considerations and so $z \approx 1$. What is very interesting is that (\ref{e10}) this time gives us the correct microwave cosmic background radiation wavelength and temperature.
\section{Renormalization}
The problem of Renormalization, as is well known was encountered first in Classical Electrodynamics \cite{rohr}. This was because the electromagnetic self energy of an electron viz., $\frac{e^2}{r}$ would $\to \infty$ as the size $r$ of the electron $\to 0$. On the other hand if $r$ were not to $\to 0$, that is the electron had a finite size, then this would lead to its own problems requiring the introduction of, for example Poincare stresses to hold the electron together (Cf.\cite{rohr}).\\
On the other hand if we were to consider the electron as having a bare mass and a physical mass, that is \cite{hooft}
\begin{equation}
m_{\mbox{phys}} = m_{\mbox{bare}} + \frac{e^2}{r}\label{e11}
\end{equation}
Then we could still have a finite physical mass, which would be what is actually measured, by allowing infinite two terms on the right side of (\ref{e11}) to cancel each other out for the limit. This means we could preserve Special Relativity and at the same time recover a finite physical mass. However in Quantum Theory, there is no real problem with Special Relativity and superluminal or non-local velocities if $r$ were to be non zero, but of the order of the Compton wavelength $l$. We can see this as follows. A particle can travel from the spacetime point $x_1$ to the spacetime point $x_2$ causally only if the interval is time like, that is
$$\eta_{\alpha \beta} (x_1 - x_2)^\alpha (x_1 - x_2)^\beta < 0$$
On the other hand because of the Uncertainty Principle there is a non zero probability for a particle to move from $x_1$ to $x_2$ even if the interval is space like, that is with superluminal velocity as long as
$$(x_1 - x_2)^2 - (x_1^0 - x_2^0)^2 \leq \frac{h^2}{m^2} (c = 1)$$
In other words there is a breakdown of causal physics within the Compton scale 
 \cite{wein}.\\
All this has direct relevance to our discussion in the previous two sections about a minimum physical scale $l$ and fuzzyness. The point is that either we invoke Quantum Mechanics and thereby recover the Compton scale, or we invoke the minimum scale $l$ and recover Quantum Mechanics (Cf. also ref.\cite{cu} for a discussion of the latter point).

\end{document}